\documentclass[10pt,aps,prl,showpacs,superscriptaddress,twocolumn,citeautoscript]{revtex4-1}

\usepackage{graphicx,color}  
\usepackage{dcolumn}   
\usepackage{bm}        
\usepackage{amssymb,url}
\usepackage{amsmath}
\usepackage{hyperref}
\usepackage{braket}

\newcommand{\beq}{\begin{equation}}
\newcommand{\eeq}{\end{equation}}
\newcommand{\ra}{\bm{r}_\alpha}

\begin{document}

\title{Acoustic Anomalous Reflectors Based on Diffraction Grating Engineering}
\author{Daniel Torrent}
\email{dtorrent@uji.es}
\affiliation{GROC, UJI, Institut de Noves Tecnologies de la Imatge (INIT), Universitat Jaume I, 12071, Castell\'o, (Spain)}
\date{\today}

\begin{abstract}
 We present an efficient method for the  design of anomalous reflectors for acoustic waves. The approach is based on the fact that the anomalous reflector is actually a diffraction grating in which the amplitude of all the modes is negligible except the one traveling towards the desired direction. A supercell of drilled holes in an acoustically rigid surface is proposed as the basic unit cell, and analytical expressions for an inverse diffraction problem are derived. It is found that the the number of holes required for the realization of an anomalous reflector is equal to the number of diffracted modes to cancel, and this number depends on the relationship between the incident and reflected angles. Then, the ``retrorreflection'' effect is obtained by just one hole per unit cell, also with only two holes it is possible to change the reflection angle of a normally incident wave and five holes are enough to design a general retroreflector changing the incident and reflected angles at oblique incidence. Finally, the concept of Snell's law violation is extended not only to the incident and reflected angles, but also to the plane in which it happens, and a device based on a single hole in a square lattice is designed in such a way that the reflection plane is rotated $\pi/4$ with respect to the plane of incidence. Numerical simulations are performed to support the predictions of the analytical expressions, and an excellent agreement is found.
 \end{abstract}

\maketitle
Anomalous reflectors and refractors can be defined as structured flat surfaces in which the relationship between the angles of the incident, reflected and refracted waves does not satisfy Snell's law\cite{born2013principles}. These devices, designed mainly in the framework of the so-called generalized laws of refraction and reflection \cite{yu2011light}, have received increasing interest within the last years \cite{zhao2013redirection,li2013reflected,li2014experimental,ma2014acoustic,xie2014wavefront,tang2014anomalous,zhai2015manipulation,li2017tunable,chen2018broadband}, and a wide variety of applications and effects have been envisioned for the control of acoustic waves, like carpet cloaks\cite{esfahlani2016acoustic}, acoustic diodes \cite{wang2016broadband}, helical wavefront generators \cite{esfahlani2017generation}, signal-processing devices \cite{zuo2017mathematical} or diffusers\cite{zhu2017ultrathin}. 

Also named ``gradient metasurfaces'', the efficient design of these devices requires of a continuous variation of the impedance of the unit cell \cite{diaz2017acoustic,li2018systematic}, which can be achieved only approximately with a large enough number of discrete resonators, with obvious practical limitations. Additionally, to achieve unitary efficiency non-local or active designs are required. The overall result is that efficient gradient metasurfaces requires a complicated design process that hinders their applications when more advanced wave-control devices are envisioned.

Recently, it has been shown that some functionalities of gradient metasurfaces for electromagnetic waves can be achieved by means of properly designed diffraction gratings based on bianostropic particles\cite{ra2017metagratings} or bipartite particles\cite{wong2018perfect,sell2018ultra}. From this perspective, the anomalous reflection or refraction effect consists essentially in cancelling all the diffracted modes except the one traveling towards the desired ``anomalous'' direction. The cancellation of all the diffracted modes except one results in the mirage that the wave has not been ``diffracted'' but ``anomalously refracted''. However, current approaches based on diffraction mode control have been applied only to retrorreflectors and anomalous reflectors at normal incidence, which will be shown here to be less demanding than the general anomalous reflector. Additionally, the acoustic counterpart of these structures has not been considered so far.

In this work we present a simplified and more general picture for the design of acoustic anomalous reflectors. The approach is based on the efficient engineering of the different diffracted modes by a periodically structured acoustic surface. The structure consists in a perforated acoustically rigid surface, and it is found that the number of holes can be set equal to the number of diffracted modes to be canceled, with the interesting result that only one or two holes are required for the most typical applications of anomalous reflectors, while only five are required for one of the most challenging applications. Finally, an off-axis anomalous reflector is designed, where the incident and reflected waves lay in different planes. 

\begin{figure}[ht!]
\begin{center}
\includegraphics[width=\linewidth]{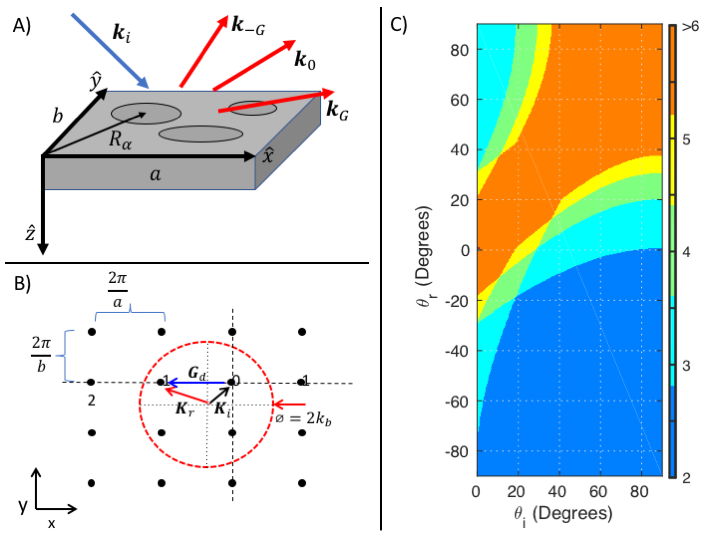}%
\caption{\label{fig:schematics} A) Schematic representation of the diffraction problem considered in the text. B) Selection of the grating geometry to generate a desired diffracted mode from a given incident plane wave with in-plane wave vectors $\bm{K}_r$ and $\bm{K}_i$, respectively. C) Number of excited diffraction orders for each incident($\theta_i$) and diffracted($\theta_r$)  angles (defined as the angle of the wave with the z-axis).}
\end{center}
\end{figure}

The proposed unit cell is shown in panel A of Fig.\ref{fig:schematics}. It consist in an acoustically rigid surface placed in the $xy$ plane at $z=0$ in which it is drilled a cluster of $N$ holes of length $L_\alpha$ and located at the positions $\ra$, for $\alpha=1,2,\ldots, N$. The cross section of the holes can be arbitrary, but it will be assumed that only the fundamental mode of the waveguide they define is excited\cite{christensen2008theory}. The holes are backed by a rigid wall, so that no energy is transferred to the other side of the surface. We assume time harmonic dependence of the fields of the form $e^{-i\omega t}$. If the surface is excited by an incident plane wave of unitary amplitud and propagating along the $z$ axis with wavenumber $\bm{k}=\bm{K}+q_0\hat{\bm{z}}$, a set of diffracted modes with reflection coefficients $R_G$ will be excited, so that the pressure and normal velocity fields will be given by
\begin{align}
P&=\sum_G\left(\delta_{G0}e^{iq_Gz}+R_Ge^{-iq_Gz}\right)e^{i\bm{K}_G\cdot\bm{r}},\\
v_n&=\frac{iq_G}{k_b Z_b}\sum_G\left(\delta_{G0}e^{iq_Gz}-R_Ge^{-iq_Gz}\right)e^{i\bm{K}_G\cdot\bm{r}},
\end{align}
with $|\bm K+\bm G|^2+q_G^2=\omega^2/c_b^2$ and with $\bm G$ being the set of all reciprocal lattice vectors. The reflectance in energy will be always unitary, but we will use the grating to engineer amount of energy that is transferred to each propagating ($Im(q_G)=0$) mode. The fields inside each hole can be set as\cite{torrent2012acoustic}
\begin{align}
P&=e^{iK\cdot R_\alpha}B_\alpha \frac{\cos k_b(z-L_\alpha)}{\sin k_bL_\alpha},\\
v_n&=-\frac{e^{iK\cdot R_\alpha}}{Z_b} B_\alpha\frac{\sin k_b(z-L_\alpha)}{\sin k_bL_\alpha},
\end{align}
which ensures the boundary condition $v_n=0$ at $z=L$. The constant factors $e^{iK\cdot R_\alpha}$ and $\sin k_bL_\alpha$ are extracted from $B_\alpha$ for later convenience.

The mode matching method is applied by projecting the Bloch modes with the $v_n$ field and the cavity modes for the $P$ field\cite{torrent2012acoustic}, resulting in the system of equations
\begin{align}
\sum_G H_{\alpha G}e^{iG\cdot R_\alpha}(\delta_{G0}+R_G)=B_\alpha \cot k_bL_\alpha, \\
\delta_{G0}-R_G=-i\frac{k_b}{q_G}\sum_\beta f_\beta H_{\beta G}e^{-iG\cdot R_\beta} B_\beta,\label{eq:vn}
\end{align}
where the coupling factor is given by $H_{\alpha G}=\frac{1}{\Omega_h}\iint_{\Omega_h}e^{iK_G\cdot(\bm{r}-\bm{R}_\alpha)}d\Omega$ and the hole's filling fraction has been defined as $f_\alpha=\frac{\Omega_\alpha}{\Omega}$, with $\Omega$ and $\Omega_\alpha$ being the areas of the unit cell and the hole $\alpha$, respectively. The above system of equations allows to solve for the $B_\alpha$ coefficients from
\beq
\label{eq:Ba}
\sum_\beta\left[\delta_{\alpha\beta}\cot k_bL_\alpha-i\chi_{\alpha\beta}\right]B_\beta=2H_{\alpha 0}
\eeq
where we have defined the interaction term $\chi_{\alpha\beta}$ as
\beq
\chi_{\alpha\beta}=\sum_G \frac{k_b}{q_G}   H_{\alpha G}H_{\beta G}f_\beta e^{-iG\cdot R_{\alpha\beta}}.
\eeq
Once the $B_\alpha$ coefficients are known, the reflection coefficient of each diffracted mode is obtained directly from equation \eqref{eq:vn}. Therefore, once the geometry of the unit cell and the holes is given, the full diffraction problem can be solved by means of the system of $N$ equations and $N$ unknowns defined by Eq. \eqref{eq:Ba}.

However, equation \eqref{eq:vn} can be used to set up an inverse problem as follows: we can impose a set of values for the amplitude of a number $g$ of diffracted modes $R_g$, design a unit cell with $N=g$ holes and solve for the $B_\alpha$ coefficients from equation \eqref{eq:vn}, since it defines a system of $g$ equations with $N=g$ unknowns. The coefficients of this system of equations are $A_{g\alpha}=f_\beta H_{\beta G}e^{-iG\cdot R_\beta}$, which depend on the position and size of the different holes but, interestingly, not on their length $L_\alpha$. Therefore, once selected the shape and position of the holes, and solved for the $B_\alpha$ coefficients, the length of each hole is directly obtained from equation $\eqref{eq:Ba}$ as
\beq
\label{eq:cotx}
\cot k_bL_\alpha=\left(2H_{\alpha 0}+i\sum_\beta\chi_{\alpha\beta}B_\beta\right)B_{\alpha}^{-1}.
\eeq

Equations \eqref{eq:vn} and \eqref{eq:cotx} constitute therefore the basis for the inverse design of diffraction gratings, however it must to be pointed out that in order to have a physically acceptable solution it is required that the right hand side of the above equation be a real number, since the $\cot x$ function is real valued for all the physically acceptable $k_b L_\alpha$ (assuming no loss or gain elements). Therefore, the additional condition 
\beq
\label{eq:Im}
\text{Im}(\cot k_bL_\alpha)=0
\eeq
has to be satisfied for a physically acceptable solution. 

The above procedure considerably simplifies the design of anomalous reflectors, in which it is desired that a wave incident with wavenumber $\bm{k}_i$ be totally reflected with wavenumber $\bm{k}_r$. From a diffraction point of view, this is equivalent to design a diffraction grating in which the reflected wave corresponds to one of the $\bm K+\bm G$ diffracted modes, and optimize the grattng in such a way that all the other propagating diffracted modes present zero amplitude. The design procedure is illustrated in panel B of figure \ref{fig:schematics}; once known the incident and reflected wavevectors we design the lattice so that their projections on the plane satisfy $\bm K_r-\bm K_i=\pm \frac{2\pi}{a}\hat{x}$.
\begin{figure}[ht!]
\begin{center}
\includegraphics[width=\linewidth]{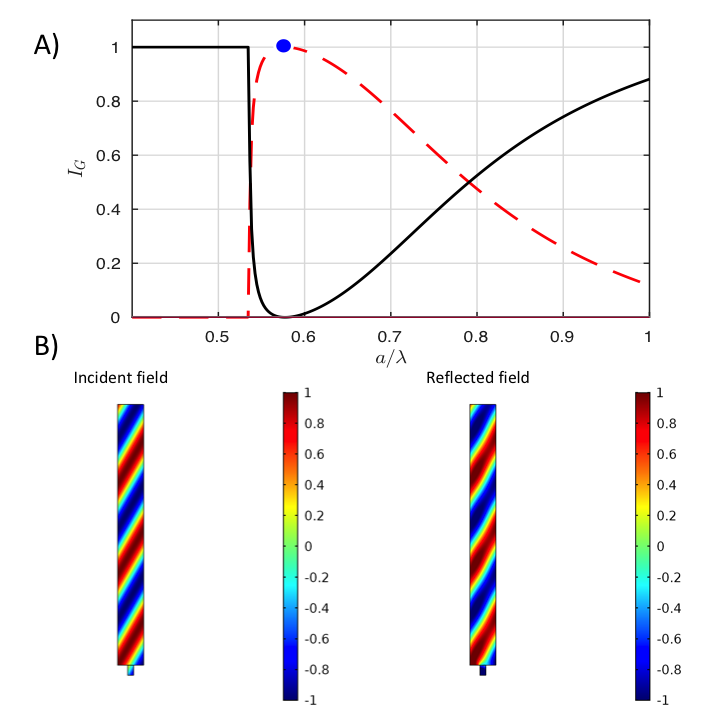}%
\caption{\label{fig:retro} A) Diffracted energy as a function of $a/\lambda$ for each propagating mode for a single groove ``retrorreflector''. B) Numerical simulation of the incident (left) and reflected (right) fields, showing the perfect retrorreflection effect.}
\end{center}
\end{figure}

The number of excited modes for a given incident and reflected waves is an intrinsic property of the lattice. Panel C of figure \ref{fig:schematics} shows this number for all the possible incident and reflected angles with the $z$ axis, $\theta_i$ and $\theta_r$, respectively, assuming in-plane reflection (although off-plane anomalous reflection is also feasible, as it will be shown in the examples below). As can be seen, the higher number of diffracted modes are excited for reflection angles similar to the incident angle, while for the ``retrorreflection'' and anomalous reflection effect at normal incidence, only one or two modes are excited and, therefore, they are less demanding devices. This interesting feature of diffraction grating is the responsible of the fact that ``extreme'' anomalous reflection be easier to implement, although the present approach offers a general method to any configuration.

The design of a ``perfect'' anomalous reflector consists in defining the incident and reflected wave vectors $\bm K_i$ and $\bm K_r$, which authomatically determines the lattice geometry and the number of excited modes $N_d$. Then we set the number of holes in the unit cell to $N=N_d-1$, since we want to impose $R_G=0$ for all the $N_d$ modes except  the one satisfying $\bm K_r=\bm K_i+\bm{G}_d$. We will then search for the size and position of the holes to satisfy condition \eqref{eq:Im} which will give us the length of the holes from Eq. \eqref{eq:cotx}. 

Four examples of application of the previous approach will be developed. In the first three the anomalous reflection effect will take place in-plane, for which a geometry invariant along the $y$ axis will be selected. In this case, the holes are grooves in the plate of width $d_\alpha$, and we have that $H^\text{groove}_{\alpha G}=\sin(|\bm K+\bm G|d_\alpha/2)/(|\bm K+\bm G|d_\alpha/2)$, while for the fourth example a cylindrical hole of radius $R_\alpha$ will be employed, and now $H^\text{hole}_{\alpha G}=2J_1(|\bm K+\bm G|R_\alpha)/(|\bm K+\bm G|R_\alpha)$.

Panel C of figure \ref{fig:schematics} shows that the retrorreflection effect can be achieved by just two diffracted modes as long as the incident angle $\theta_i=-\theta_r$ be higher than approximately $20^\circ$ (dark blue region), threrefore we will need only one hole per unit cell to design such a device. The condition $R_0=0$  in equation \eqref{eq:vn} we get $B_0=\frac{iq_0}{f_0H_0k_b}$ and inserting this into equation \eqref{eq:cotx} and setting the imaginary part of $\cot k_bL_0$ equal to zero we get the condition for energy conservation,
\beq
\label{eq:condretro}
\frac{q_{G_d}H_0^2}{q_0H_{G_d}^2}=1,
\eeq
what give us
\beq
\label{eq:cotretro}
\cot k_bL_0=\text{Im}({\chi_{00}}).
\eeq

Given that in condition \ref{eq:condretro} the functions $H_G$ and $q_G$ are computed at $\bm K_i$($\bm G=0$) and $\bm K_r$ ($\bm G=\bm G_d$), this condition is trivially satisfied when $\bm K_r=-\bm K_i$, therefore the design method consists in selecting the width $d_0$ of the groove and obtaining $L_0$ from equation \ref{eq:cotretro}. In our first example we select $\theta_i=\pi/3$, so that the retrorreflection diffraction condition is satisfied at $a/\lambda=2\sin\theta_i=1.73$, selecting $d_0=0.23a$ defines $L_0=0.23a$. 

Figure \ref{fig:retro}, panel A) shows the diffraction energy $I_G=q_G/q_0|R_G|^2$ as a function of $a/\lambda$ for the designed retrorreflector. We see how the energy reflected by the fundamental mode (black line) becomes zero and all the energy goes to the first diffraction order (red-dashed line) at the desired $a/\lambda$ point. Panel B) shows the numerical simulations performed with the comercial finite element sotware COMSOL Multiphysics, verifying that the incident (left) and reflected (right) waves have the same propagation direction. 

The second example analyzed is  the anomalous reflector at normal incidence, in which a wave incides normally to the surface and it is reflected an angle $\theta_r$. In this case, for  desired reflection angles higher than $\pi/6$ we have only three diffracted modes, the fundamental one and the lateral ones at $\pm \theta_r$, the objective is to cancel the fundamental and one of the diffracted orders, so that we need only two grooves per unit cell. There are obviously a large number of degrees of freedom, but we will propose a unit cell in which the two grooves, labeled $\alpha$ and $\beta$, are identical and symmetrically placed in the unit cell, $x_\beta=-x_\alpha$. The condition $R_0=R_G=0$ gives now $B_\alpha=\frac{iq_0}{k_bf_0H_0}\frac{1}{1-e^{iGx_{\alpha\beta}}}$, where $x_{\alpha\beta}=x_\beta-x_\alpha=-2x_\alpha$.

\begin{figure}[ht!]
\begin{center}
\includegraphics[width=\linewidth]{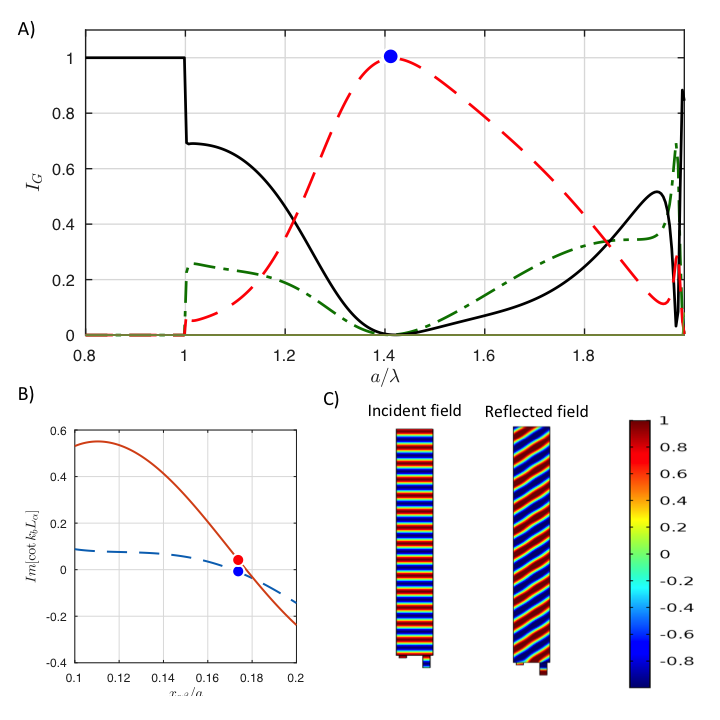}%
\caption{\label{fig:AR} A) Diffracted energy as a function of $a/\lambda$ for each propagating mode for a two-grooves anomalous reflector at normal incidence. C) Plot of $\text{Im}(\cot k_b L_\alpha)$ as a function of the groove's semi-distance $x_{\alpha\beta}/a$, showing the points that satisfy the energy balance condition at $x_{\alpha\beta}=0.32a$. C) Numerical simulation of the incident (left) and reflected (right) fields.}
\end{center}
\end{figure}

Figure \ref{fig:AR}, panel A) shows the diffracted energy $I_G$ in this example, where we have selected $\theta_r=\pi/4$, which sets $\lambda/a=0.7071$. It is clear how the energy of the fundamental (red line) and one diffracted (green dot-dashed line) modes cancel at the desired wavelength. The width of the grooves is set as $d_0=0.2a$, and the distance between them is obtained from condition \eqref{eq:Im}, which is plotted in panel B) of figure \ref{fig:AR} as a function of $x_\alpha/a$. Finally, the incident and reflected fields computed with COMSOL are depicted in panel C) of the figure.

\begin{figure}[ht!]
\begin{center}
\includegraphics[width=\linewidth]{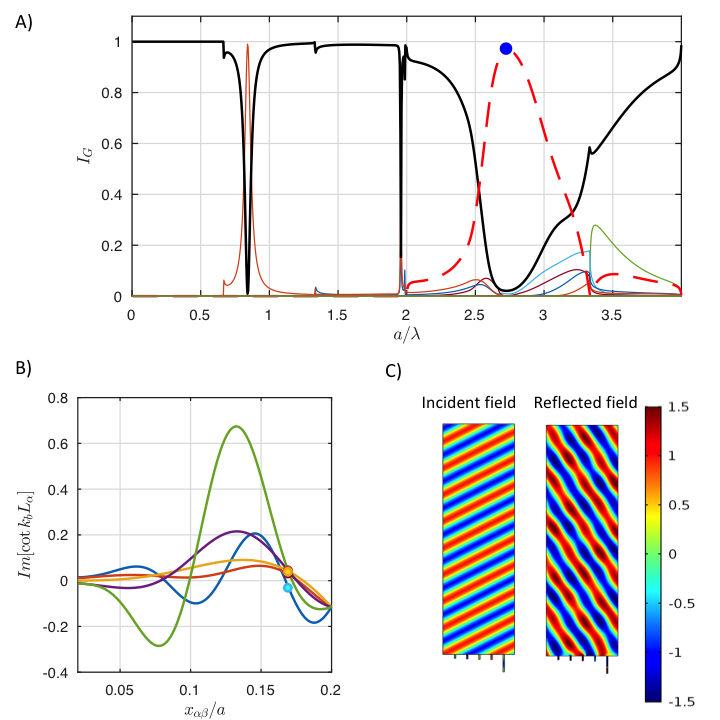}%
\caption{\label{fig:ARangle} A) Diffracted energy as a function of $a/\lambda$ for each propagating mode for a five-grooves anomalous reflector. B) Plot of $\text{Im}(\cot k_b L_\alpha)$ as a function of the groove's semi-distance $x_{\alpha\beta}/a$, showing the points that satisfy the energy balance condition at $x_{\alpha\beta}=0.17a$. C) Numerical simulation of the incident (left) and reflected (right) fields.}
\end{center}
\end{figure}

Next we show an example of an anomalous reflector, in which the reflection angle of the wave is changed but keeping the same sign. We select $\theta_i=\pi/3$ and $\theta_r=\pi/6$, which corresponds to $N_d=6$ in panel C) of figure \ref{fig:schematics}, therefore this interesting effect can be obtained with only $N=5$ grooves. We set the size of the hole as $d_0=0.02a$ and the $B_\alpha$ coefficients are directly obtained from the solution of the system of equations defined by Eq. \eqref{eq:vn}. The result of the design can be found in the plot of the diffracted energy in figure \ref{fig:ARangle}, panel A), where the distance between grooves $x_{\alpha\beta}=0.17a$ that minimizes the imaginary part of $\cot k_bL_\alpha$ has been obtained from the plot of panel B) as in the previous example, and the required lengths of the grooves are $L_\alpha=0.0613a,0.0715a,0.0776a,0.0835a$ and $0.2775a$. Panel C) shows the incident $P_0$ and reflected $P_S$ waves as simulated with COMSOL, illustrating the nearly perfect performance of the device. It has to be pointed out that the optimal distance between grooves is not $a/N$, so that actually the cluster of grooves does not form a sub-lattice of the main lattice, as it happens in devices based on phase gradients.

\begin{figure}[ht!]
\begin{center}
\includegraphics[width=\linewidth]{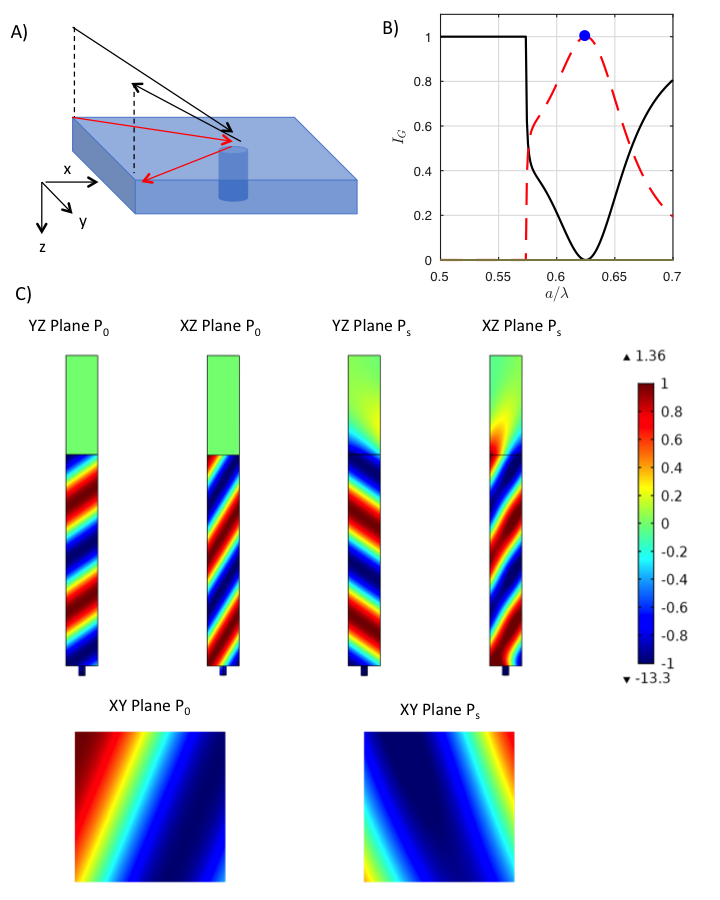}%
\caption{\label{fig:3Dretro} A) Geometry of the off-axis reflection problem. Diffracted energy as a function of $a/\lambda$ for each propagating mode for a reflector made of a single circular hole in a square unit cell. C) Projections of the incident and reflected fields at the different planes of the unit cell.}
\end{center}
\end{figure}

Finally, the presented theory is applied to the design of a four-channel off-axis reflector. This device reflects the incident wave backwards but rotated a given angle in the xy plane, as illustrated in panel A) of figure \ref{fig:3Dretro}, therefore the incident and reflected waves are in two different planes, in contradiction with Snell's law. We select the incident and reflected angles with the z axis of $\theta_i=-\theta_r=\pi/3 $ and the rotation angle $\theta_t=\pi/4$, only one hole per unit cell is required, and selecting circular holes in a square lattice ensures a four-channel functionality, due to the four-fold symmetry of this lattice. The design method is identical as to the retrorreflector of figure \ref{fig:retro}, since equation \eqref{eq:condretro} is trivially satisfied (the projection of the wavenumber remains unchanged) and the length of the cylindrical hole is obtained from equation \eqref{eq:cotretro}. Figure \ref{fig:3Dretro} panel B) shows the diffracted energy and panel C) shows the numerical simulations performed with COMSOL of the incident and reflected fields projected at the different sides of the three-dimensional unit cell, showing the retrorreflection effect responsable of the rotation of the reflection angle. It is remarkable the simplicity of this device as compared with the equivalent gradient-phase metasurface that would be required for this functionality.

In summary we have shown that anomalous reflection from acoustic surfaces can be properly and efficiently obtained by means of engineered diffraction gratings, in which subwavelength holes are drilled in an acoustically rigid surface. The number of holes required is in general one less than the number of diffracted modes, so that all these modes are cancelled except one, which is the carrier of the wave at the desired reflection angle. It has been shown that unitary efficiency can be achieved for one and two holes per unit cell, and nearly unitary in the case of five holes, showing also the great potential that this method has for the design of more efficient anomalous reflectors. This approach presents several advantages in comparison with previous approaches based on gradient index metasurfaces, since no continuous variation of the index or the surface's is required, but just a discrete number of properly selected holes. The presented theory therefore opens the door to a new set of devices efficiently designed for the full control of the propagation direction of acoustic waves. Finally, this approach could be applied as well to anomalous refractors and to other domains of physics, like elasticity or electromagnetism, since the principles in which it is based are general for all type of waves. 


\begin{acknowledgments}
Work supported by the LabEx AMADEus (ANR-10-LABX-42) in the framework of IdEx Bordeaux
(ANR-10-IDEX-03-02) and by the U.S. Office of Naval Research under Grant No. N00014-17-1-2445. D.T. acknowledges financial support through the ``Ram\'on y Cajal'' fellowship.
\end{acknowledgments}

%
%
%
%

\end{document}